\documentclass[11pt]{article}
\usepackage[margin=1in]{geometry}
\usepackage{amsmath,amssymb}
\usepackage{graphicx}
\usepackage{booktabs}
\usepackage{tikz}
\usetikzlibrary{arrows.meta,positioning,shapes.geometric,fit,backgrounds}
\usepackage{hyperref}
\hypersetup{colorlinks=true,linkcolor=blue,citecolor=blue,urlcolor=blue}
\usepackage[numbers,sort&compress]{natbib}
\usepackage{caption}
\usepackage{xcolor}
\usepackage{enumitem}
\usepackage{authblk}

\title{\Large A Unified Multi-Modal Sensing and Active-Stabilization Framework for Autonomous IoT Nodes in Connectivity-Denied Environments: From Perimeter Sentinel to High-Value Cargo Protection}

\author[1]{Naahi Mumtaj Rihan}
\affil[1]{Department of Electrical and Electronic Engineering, BRAC University, Dhaka, Bangladesh}
\date{\today}

\begin{document}
\maketitle

\begin{abstract}
\noindent
Autonomous embedded nodes deployed in connectivity-denied environments -- remote forest perimeters, farm boundaries, and cargo in transit across highways or open ocean -- share a common set of engineering requirements that are usually addressed by separate, purpose-built systems: passive-infrared (PIR) triggered wake-up, inertial-measurement-unit (IMU) driven active stabilization, and long-range communication under strict power budgets. This paper argues that a perimeter/wildlife security turret and a high-value cargo protection crate are, from a systems perspective, two parametrizations of the \emph{same} generalized node architecture rather than two unrelated designs. We formalize this generalized architecture, extend it with three sensing/communication modalities not present in either original design -- single-point LiDAR ranging, a LoRa/LoRaWAN regional mesh tier, and a satellite short-burst-data (SBD) global tier -- and couple the resulting telemetry to a Geographic Information System (GIS) layer via a proposed Composite Risk Index (CRI) that fuses breach events, impact severity, and connectivity state into a single spatially-referenced score. We present the sensor-fusion, proportional-integral-derivative (PID) stabilization, link-budget, received-signal-strength-indicator (RSSI) localization, and CRI formulations that unify the two deployment modes, provide a simulated closed-loop stabilization response and a tiered-communication trade-off analysis, and instantiate the two original projects as case studies of the generalized architecture. The framework is intended as an engineering blueprint and a starting point for field validation rather than as a report of field-tested results.
\end{abstract}


\section{Introduction}
\label{sec:intro}

\subsection{Motivation}
Two classes of embedded systems are conventionally treated as unrelated: (i) static \emph{perimeter/wildlife sentinel} nodes that wake on motion and aim a camera or deterrent at a target, and (ii) mobile \emph{cargo protection} platforms that actively counteract shocks to protect a fragile payload in transit. Both, however, are built from the same primitive: a low-power microcontroller that (a) wakes on an event detected by a passive sensor, (b) reads an inertial or ranging sensor to characterize a physical disturbance, (c) drives an actuator through a closed feedback loop to correct for that disturbance or to track a target, and (d) reports the event over a communication link that cannot assume the presence of local Wi-Fi or a wired network. This paper's central claim is that treating these as instances of a single generalized architecture -- rather than as two point designs -- exposes design decisions (sensing modality, communication tier, control gain scheduling) that are obscured when each system is engineered in isolation.

\subsection{Contributions}
This work makes the following contributions:
\begin{enumerate}[leftmargin=*]
    \item A generalized node architecture in which the perimeter sentinel and cargo-protection systems are recovered as two operating-mode instantiations of one hardware/firmware template (Section~\ref{sec:arch}).
    \item Extension of the sensing suite with single-point LiDAR time-of-flight ranging, formalized alongside the existing PIR and IMU sensors through a lightweight complementary/Kalman fusion model (Section~\ref{sec:fusion}).
    \item A closed-loop PID stabilization formulation with anti-windup, applied identically to line-of-sight tracking (sentinel mode) and payload counter-tilt (cargo mode), together with a simulated step/disturbance response (Section~\ref{sec:pid}).
    \item An adaptive, three-tier communication stack (local RF/RSSI, cellular GSM/GPRS, regional LoRa/LoRaWAN, global satellite SBD) with link-budget and duty-cycle equations governing tier selection (Section~\ref{sec:comm}).
    \item A proposed Composite Risk Index that fuses breach events, impact severity, and connectivity state onto a GIS layer for both deployment modes (Section~\ref{sec:gis}).
    \item Two worked case studies recovering the original perimeter-sentinel and cargo-protection designs as parameter settings of the unified framework (Section~\ref{sec:case}).
\end{enumerate}

\subsection{Scope and Positioning}
This paper is a systems/architecture contribution. The PID, sensor-fusion, and link-budget equations are standard control-theory and communications formulations applied to a new combined use case; the Composite Risk Index (Section~\ref{sec:gis}) is the paper's proposed novel element. All figures showing quantitative behavior (Figs.~\ref{fig:pid},~\ref{fig:comm},~\ref{fig:gis}) are simulated or synthetic illustrations of the model, clearly distinguished from field-measured data, since no hardware instantiation of the fully unified system has yet been built.

\section{Related Work}
\label{sec:related}

\textbf{Gimbal and line-of-sight stabilization.} Inertially stabilized platforms have a long history in electro-optical and missile-seeker applications, with the foundational control formulation and equations of motion given by Hilkert~\citep{hilkert2008} and Ekstrand~\citep{ekstrand2001}. PID and PID-variant controllers remain the dominant approach for two- and three-axis camera/sensor gimbals because of their low computational cost relative to fuzzy, sliding-mode, or neural-network alternatives, which the present work exploits for its ESP32-class target hardware. Closest in spirit to the sensing side of this paper is the work of Assaf~et~al.~\citep{assaf2022}, who couple a high-precision, low-cost pointing mechanism to a narrow-field-of-view 1D LiDAR for long-range railway obstacle detection -- directly motivating the PIR-plus-LiDAR sensing pattern adopted in Section~\ref{sec:fusion}.

\textbf{Low-power wide-area and satellite communication.} LoRa/LoRaWAN is well established as a low-power, long-range alternative or complement to cellular connectivity for IoT nodes; Haxhibeqiri~et~al.~\citep{haxhibeqiri2018} and Raza~et~al.~\citep{raza2017} survey its physical- and network-layer characteristics, and Shanmuga Sundaram~et~al.~\citep{shanmuga2020} catalogue open research problems in large-scale LoRa deployments. For coverage beyond cellular and LoRa gateway range, satellite short-burst-data services are the standard solution for globally mobile assets; Ferrer~et~al.~\citep{ferrer2019} review medium-access-control protocols for nanosatellite-based IoT, and the Iridium SBD service~\citep{iridium2026} is used in this paper as a representative global tier because of its established use in maritime and logistics tracking.

\textbf{Cargo shock/vibration monitoring.} Commercial shock and vibration data loggers are widely used to document handling conditions for fragile or regulated shipments; Meng and Zhu~\citep{meng2020} demonstrate a low-cost IoT vibration-sensing architecture (MEMS accelerometer, cellular backhaul, cloud alerting) for a related civil-engineering monitoring problem, whose data pipeline this paper adapts for in-transit shock telemetry.

\textbf{Perimeter intrusion sensing.} 3D LiDAR-based perimeter intrusion detection is reported by industrial vendors to substantially reduce false-alarm rates relative to PIR-only or video-only systems~\citep{quanergy2026}, motivating this paper's use of a low-cost single-point LiDAR as a confirmatory second sensing modality after a PIR wake event, rather than as a replacement for PIR.

\textbf{RF-based localization.} Received-signal-strength-indicator (RSSI) techniques remain the most widely deployed range-estimation method in resource-constrained wireless sensor networks due to their negligible hardware cost; Du~et~al.~\citep{du2022} propose a multilateration refinement (the weighted-three-minimum-distances method) that improves on raw log-distance path-loss estimates, which informs the direction/likely-approach estimation discussed in Section~\ref{sec:comm}.

\textbf{Geospatial risk fusion in logistics.} Spatial risk mapping is an established practice in supply-chain and logistics GIS, and recent work has begun coupling real-time IoT sensor streams to spatiotemporal graph-learning models for dynamic routing; Xue~et~al.~\citep{xue2026} propose a Risk-Aware Dynamic Routing framework that predicts congestion risk from IoT and GPS data to drive routing decisions. The Composite Risk Index proposed in Section~\ref{sec:gis} is complementary to this line of work: rather than predicting traffic congestion, it fuses breach/impact telemetry from the node itself with static geospatial context to produce a per-location threat/risk score.

\section{Unified System Architecture}
\label{sec:arch}

\subsection{One Node, Two Operating Modes}
\begin{figure}[htbp]
\centering
\begin{tikzpicture}[
    node distance=8mm and 10mm,
    box/.style={draw, rounded corners, minimum height=0.9cm, minimum width=2.6cm, align=center, font=\small, fill=blue!5},
    mcu/.style={draw, rounded corners, minimum height=2.6cm, minimum width=3.0cm, align=center, font=\small, fill=blue!12, line width=0.9pt},
    lbl/.style={font=\scriptsize, align=center},
    >={Latex[length=2mm]}
]
\node[mcu] (mcu) {ESP32\\(dual-core)\\PID + State\\Machine +\\Mode Flag};

\node[box, above left=4mm and 6mm of mcu] (pir) {PIR\\(wake trigger)};
\node[box, left=6mm of mcu] (imu) {IMU\\(MPU6050/9250)};
\node[box, below left=4mm and 6mm of mcu] (lidar) {LiDAR\\(1D ToF ranging)};

\node[box, above right=4mm and 6mm of mcu] (servo) {Servo(s)\\Aim / Counter-tilt};
\node[box, right=6mm of mcu] (rf) {RF/RSSI\\local sensing};
\node[box, below right=4mm and 6mm of mcu] (comm) {Comm. Stack\\GSM$\to$LoRa$\to$SBD};

\node[box, below=10mm of mcu, fill=orange!12] (power) {Battery + Buck\\Converters};
\node[box, above=10mm of mcu, fill=green!8] (gis) {Cloud / GIS\\Composite Risk\\Index};

\draw[->] (pir) -- (mcu);
\draw[->] (imu) -- (mcu);
\draw[->] (lidar) -- (mcu);
\draw[->] (mcu) -- (servo);
\draw[<->] (mcu) -- (rf);
\draw[->] (mcu) -- (comm);
\draw[->] (comm) -- (gis);
\draw[->] (power) -- (mcu);
\draw[->] (power) -- (servo);
\draw[->] (power) -- (comm);

\node[lbl, below=1mm of mcu, yshift=-16mm] {};
\end{tikzpicture}
\caption{Generalized node architecture shared by both deployment modes. Sensing (PIR, IMU, LiDAR), actuation (servo), local RF sensing, and the tiered communication stack all attach to a common ESP32 core; only the \texttt{MODE} flag, setpoint definition, and mechanical mounting differ between Sentinel and Cargo instantiations.}
\label{fig:arch}
\end{figure}
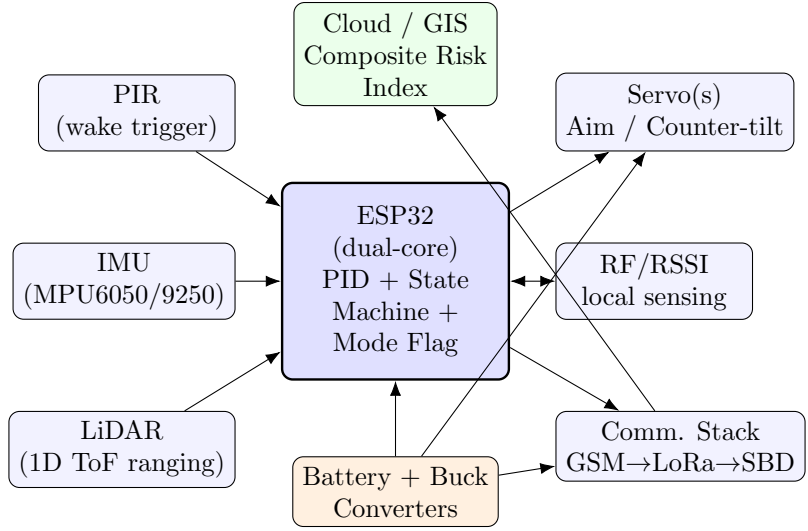

We define a single hardware/firmware template with a configuration flag \texttt{MODE} $\in \{\textsc{Sentinel}, \textsc{Cargo}\}$ that changes (i) the physical mounting and interpretation of the actuator (aiming a camera/deterrent vs. leveling a payload tray), (ii) the setpoint of the control loop (track a moving target vs. hold zero tilt), and (iii) the default communication tier ordering (cellular-first for a fixed perimeter node vs. satellite-aware for a node that may cross ocean segments). All other subsystems -- PIR wake logic, IMU/LiDAR fusion, PID core, and the communication/GIS stack -- are shared. Figure~\ref{fig:arch} shows the generalized block diagram.

\subsection{Sensing Suite}
Three sensing modalities cooperate in both modes:
\begin{itemize}[leftmargin=*]
    \item \textbf{PIR} provides a near-zero-power, always-armed wake/breach trigger (Sentinel: exterior motion; Cargo: interior breach).
    \item \textbf{IMU} (6- or 9-axis) measures the disturbance the control loop must reject (Sentinel: wind-induced pole/branch sway; Cargo: impact-induced tilt/shock).
    \item \textbf{LiDAR} (single-point time-of-flight) is added in this unified design as a confirmatory, higher-resolution modality: in Sentinel mode it range-gates the PIR trigger against a known background distance to reduce false alarms from foliage motion; in Cargo mode it can monitor internal clearance between the payload and crate wall as an independent check on the mechanical stops (Section~\ref{sec:fusion}).
\end{itemize}

\subsection{Actuation}
A servo-driven platform executes the control law of Section~\ref{sec:pid}: in Sentinel mode it re-aims a camera/spotlight/deterrent along up to two axes; in Cargo mode it counter-tilts a payload tray, with physical travel limits enforced identically in both modes to bound worst-case actuator authority.

\subsection{Communication and Geospatial Layer}
Both modes share the tiered communication stack of Section~\ref{sec:comm} and report into the same GIS/CRI backend of Section~\ref{sec:gis}; the only difference is the default tier-selection policy (Section~\ref{sec:comm}).

\section{Sensor Fusion and Ranging Model}
\label{sec:fusion}

\subsection{LiDAR Ranging}
For a single-point time-of-flight LiDAR, distance $d$ is recovered from the round-trip time of flight $\tau$ of an emitted light pulse:
\begin{equation}
d = \frac{c\,\tau}{2}
\label{eq:tof}
\end{equation}
where $c$ is the speed of light. For a sensor with angular beam divergence $\phi$, the transverse spatial resolution at range $d$ is
\begin{equation}
\Delta y \approx 2\,d\,\tan\!\left(\frac{\phi}{2}\right).
\label{eq:angres}
\end{equation}
Equation~\eqref{eq:angres} sets the practical detection-cone size used to range-gate PIR triggers in Sentinel mode and to monitor internal clearance in Cargo mode.

\subsection{Complementary/Kalman Fusion of IMU and LiDAR}
The platform's angular state estimate $\hat\theta_k$ at discrete time step $k$ combines a high-rate, drift-prone gyroscope integration with a lower-rate, drift-free reference derived from LiDAR-confirmed geometry (e.g., a known mounting reference distance). A standard discrete Kalman predict/update pair is used:
\begin{align}
\text{Predict:}\quad & \hat\theta_{k}^{-} = \hat\theta_{k-1} + \omega_{k-1}\,\Delta t, \qquad P_{k}^{-} = P_{k-1} + Q \label{eq:kalman_predict}\\
\text{Update:}\quad & K_k = \frac{P_k^{-}}{P_k^{-} + R}, \qquad \hat\theta_k = \hat\theta_k^{-} + K_k\left(z_k - \hat\theta_k^{-}\right), \qquad P_k = (1-K_k)P_k^{-} \label{eq:kalman_update}
\end{align}
where $\omega_{k-1}$ is the gyro rate, $z_k$ is the accelerometer- or LiDAR-geometry-derived angle observation, $Q$ is process noise (gyro drift), and $R$ is measurement noise. This fused estimate $\hat\theta_k$, rather than raw gyro integration, is the error signal fed to the PID loop of Section~\ref{sec:pid}, consistent with the Kalman-filtered IMU fusion reported in prior camera-gimbal work~\citep{hilkert2008}.

\section{Active Stabilization Control}
\label{sec:pid}

\subsection{PID Formulation}
The control task, running as a dedicated high-priority FreeRTOS task on one ESP32 core, computes the actuator command $u(t)$ from the fused angular error $e(t) = \theta_{\text{set}} - \hat\theta(t)$:
\begin{equation}
u(t) = K_p\,e(t) + K_i \int_0^t e(\tau)\,d\tau + K_d\,\frac{de(t)}{dt}.
\label{eq:pid}
\end{equation}
The integral term is clamped (anti-windup) to the platform's mechanical travel limit $\theta_{\max}$:
\begin{equation}
\int_0^t e(\tau)\,d\tau \;\leftarrow\; \operatorname{clip}\!\left(\int_0^t e(\tau)\,d\tau,\; -\theta_{\max}/K_i,\; \theta_{\max}/K_i\right).
\label{eq:antiwindup}
\end{equation}

\subsection{Closed-Loop Model}
Approximating the servo-driven platform as a rotational inertia $J$ with viscous damping $b$, the open-loop plant transfer function is
\begin{equation}
G(s) = \frac{1}{Js^2 + bs}.
\label{eq:plant}
\end{equation}
With the PID controller $C(s) = K_p + K_i/s + K_ds$, the closed-loop transfer function from setpoint $\Theta_{\text{set}}(s)$ to platform angle $\Theta(s)$ is
\begin{equation}
\frac{\Theta(s)}{\Theta_{\text{set}}(s)} = \frac{C(s)G(s)}{1 + C(s)G(s)} = \frac{K_d s^2 + K_p s + K_i}{Js^3 + (b + K_d)s^2 + K_p s + K_i}.
\label{eq:closedloop}
\end{equation}
Equation~\eqref{eq:closedloop} is a standard third-order closed-loop form; $K_p$, $K_i$, $K_d$ are tuned (e.g., via Ziegler--Nichols or Cohen--Coon initialization, then refined empirically) so that all closed-loop poles lie in the open left half-plane with adequate damping margin, matching the tuning practice reported across the gimbal-stabilization literature~\citep{ekstrand2001,assaf2022}.

\subsection{Simulated Response}
\begin{figure}[htbp]
\centering
\includegraphics[width=0.85\textwidth]{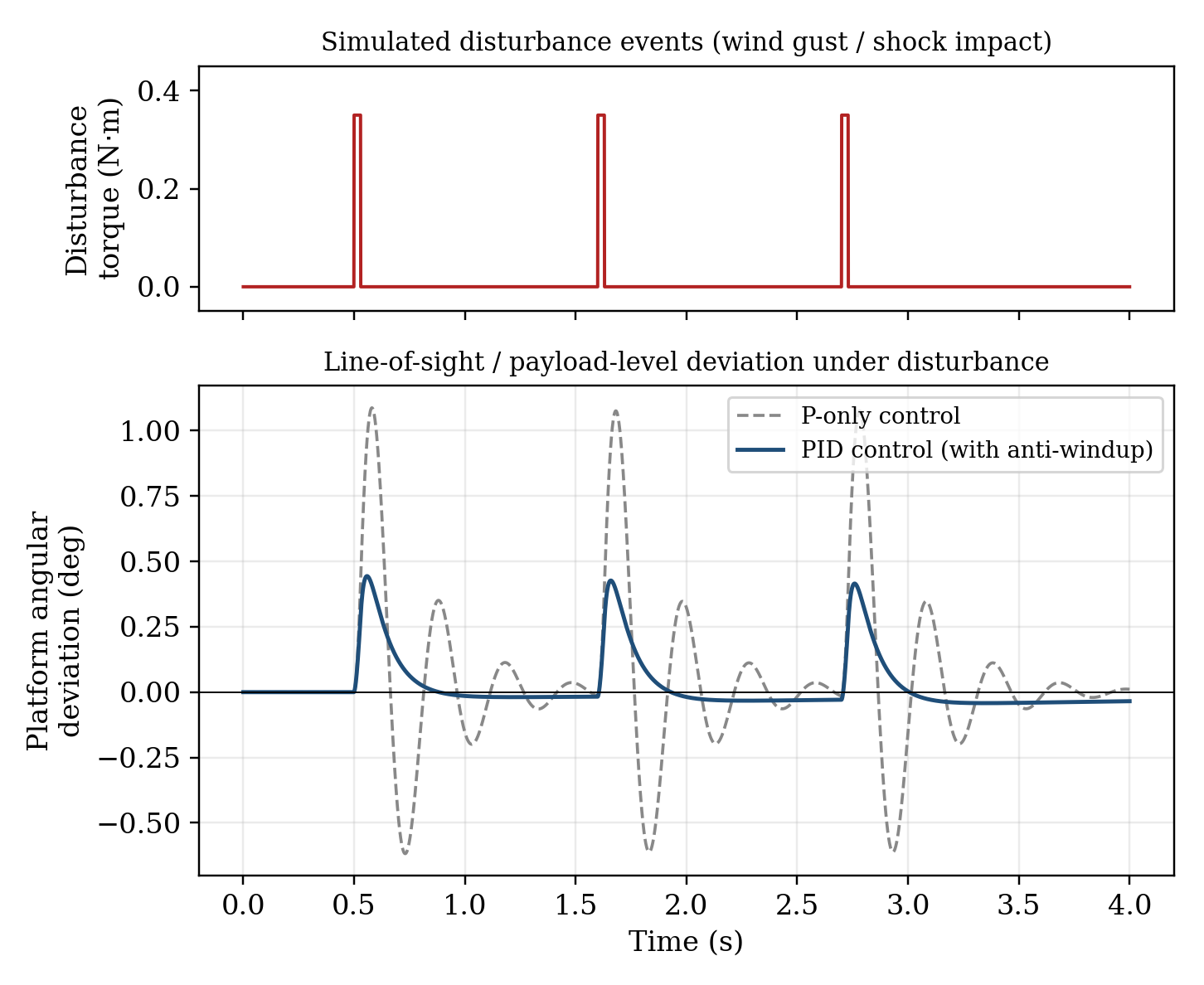}
\caption{Simulated closed-loop response of Eq.~\eqref{eq:pid}--\eqref{eq:plant} to three disturbance pulses (top), comparing full PID control against a proportional-only baseline (bottom). Parameters: $J=0.02\,\mathrm{kg\,m^2}$, $b=0.15\,\mathrm{N\,m\,s/rad}$, $K_p=9.0$, $K_i=3.0$, $K_d=0.9$.}
\label{fig:pid}
\end{figure}

Figure~\ref{fig:pid} shows a discrete-time simulation of Eq.~\eqref{eq:pid}--\eqref{eq:antiwindup} against Eq.~\eqref{eq:plant}, subjected to three short disturbance-torque pulses representing wind gusts (Sentinel mode) or shock impacts (Cargo mode). The PID controller settles the platform angle back toward the setpoint substantially faster and with less residual deviation than a proportional-only controller under the same disturbance sequence; this is a simulation of the model, not a measurement from built hardware.

\section{Adaptive Multi-Tier Communication}
\label{sec:comm}

\subsection{Tier Selection Logic}
The unified design adds two communication tiers absent from the original single-tier (GSM-only) designs: a LoRa/LoRaWAN regional mesh and a satellite SBD global tier. The firmware selects the lowest-cost tier that is currently reachable, escalating outward:
\begin{equation}
\text{Tier} =
\begin{cases}
\text{RF/RSSI (local, node-to-node)} & \text{if peer node within range,}\\
\text{GSM/GPRS} & \text{if cellular RSSI} \geq \tau_{\text{GSM}},\\
\text{LoRa/LoRaWAN} & \text{if cellular unavailable and gateway reachable,}\\
\text{Satellite SBD} & \text{otherwise (global fallback).}
\end{cases}
\label{eq:tierselect}
\end{equation}
This ordering favors the tier with the lowest energy cost per successfully delivered message, consistent with LPWAN energy-efficiency findings~\citep{haxhibeqiri2018,raza2017}, while guaranteeing that a breach or impact alert is never silently dropped for lack of cellular coverage -- the central limitation of the original single-tier cargo design.

\subsection{Link Budget}
For the terrestrial tiers (GSM and LoRa), free-space path loss between node and gateway/tower is
\begin{equation}
L_{\text{fs}}(\text{dB}) = 20\log_{10}(d) + 20\log_{10}(f) + 32.44,
\label{eq:friis}
\end{equation}
with $d$ in km and $f$ in MHz. The received signal margin available for a link to close is
\begin{equation}
M = P_{tx} + G_{tx} + G_{rx} - L_{\text{fs}} - S_{\text{rx}},
\label{eq:margin}
\end{equation}
where $S_{\text{rx}}$ is the receiver sensitivity. LoRa's chirp-spread-spectrum modulation achieves substantially lower $S_{\text{rx}}$ (higher sensitivity) than GSM at the cost of data rate, which is what permits its longer range at comparable transmit power~\citep{haxhibeqiri2018,shanmuga2020}. For a LoRa packet, the airtime (and hence energy cost) is governed by the symbol duration $T_s = 2^{SF}/BW$ and the resulting time-on-air
\begin{equation}
T_{\text{packet}} = T_{\text{preamble}} + n_{\text{sym}} \cdot T_s,
\label{eq:lora_airtime}
\end{equation}
where $SF$ is the spreading factor and $BW$ the channel bandwidth; higher $SF$ improves range/sensitivity but increases $T_{\text{packet}}$ and therefore energy per message, motivating an adaptive-data-rate policy for the tiered stack in Eq.~\eqref{eq:tierselect}.

\subsection{RSSI-Based Local Direction Estimation}
For the local RF/RSSI tier (used for node-to-node mesh awareness and coarse intruder-direction estimation in Sentinel mode), the log-distance path-loss model relates received power to distance $d$:
\begin{equation}
\text{RSSI}(d) = \text{RSSI}(d_0) - 10\,n\,\log_{10}\!\left(\frac{d}{d_0}\right) + X_\sigma,
\label{eq:pathloss}
\end{equation}
where $d_0$ is a reference distance, $n$ is the path-loss exponent, and $X_\sigma$ is log-normal shadowing noise. Because Eq.~\eqref{eq:pathloss} alone is noise-sensitive, we follow the weighted-multilateration refinement of Du~et~al.~\citep{du2022}, which combines multiple anchor-relative distance estimates with a weighted-least-squares position solve rather than trusting any single RSSI-derived range.

\begin{figure}[htbp]
\centering
\includegraphics[width=0.82\textwidth]{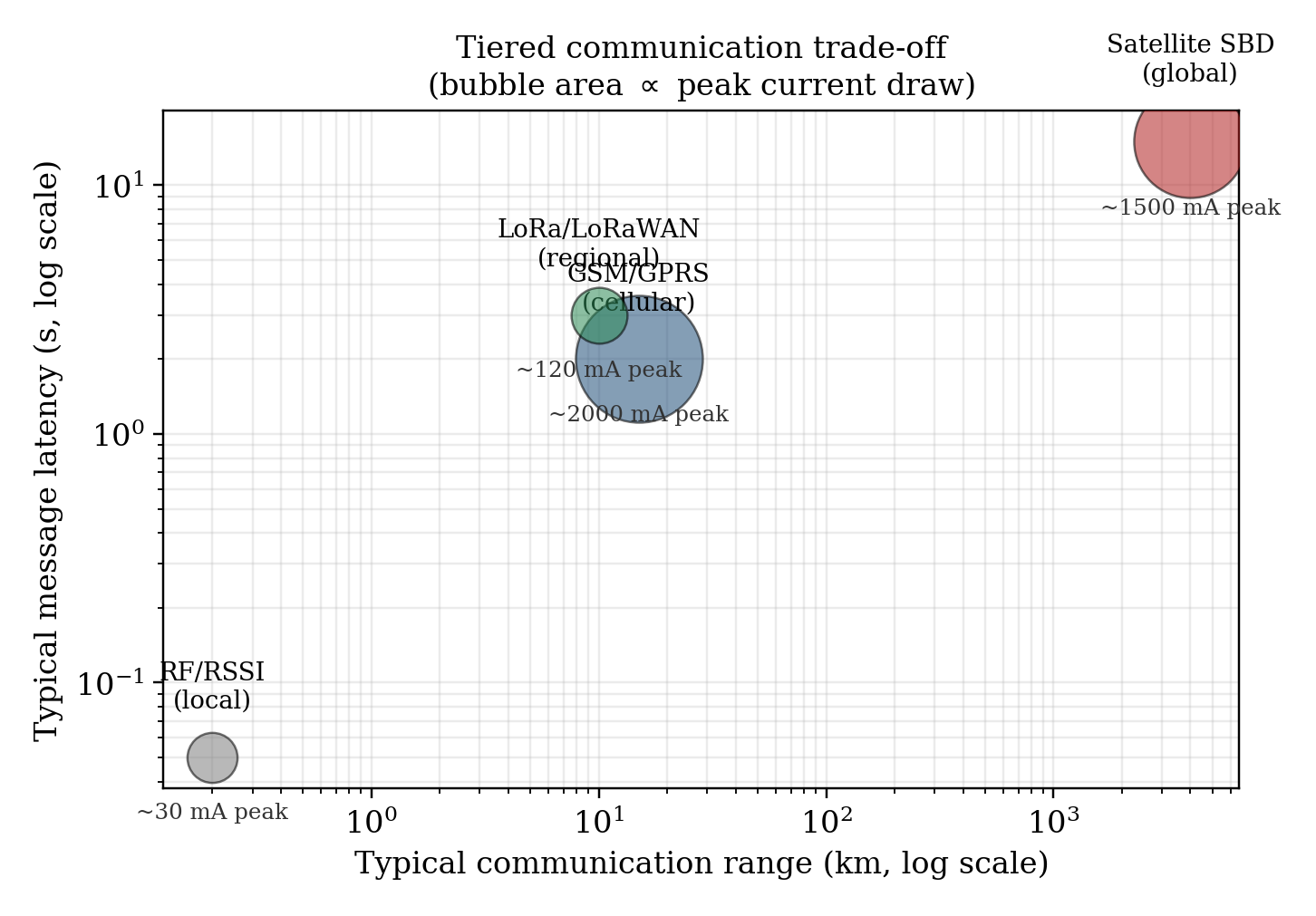}
\caption{Illustrative range/latency/power trade-off across the four communication tiers used in the unified architecture (representative, literature-informed orders of magnitude; not a measured benchmark). Bubble area is proportional to typical peak current draw, motivating why the tier-selection policy of Eq.~\eqref{eq:tierselect} favors lower tiers whenever reachable.}
\label{fig:comm}
\end{figure}

\section{Geospatial Risk Fusion}
\label{sec:gis}

\subsection{Composite Risk Index}
We propose a Composite Risk Index (CRI) $R_i \in [0,1]$ for a given node or route waypoint $i$, combining live telemetry with static geospatial context:
\begin{equation}
R_i = w_1 \hat{B}_i + w_2 \hat{I}_i + w_3 \hat{C}_i + w_4 \hat{G}_i, \qquad \sum_{k=1}^{4} w_k = 1,
\label{eq:cri}
\end{equation}
where
\begin{itemize}[leftmargin=*]
    \item $\hat{B}_i$ is a normalized breach-event score (recent PIR/LiDAR-confirmed intrusion frequency at location $i$),
    \item $\hat{I}_i$ is a normalized impact-severity score (recent peak IMU-measured g-force at location $i$),
    \item $\hat{C}_i$ is a normalized \emph{connectivity-risk} score, $\hat{C}_i = 1 - \bar{M}_i / M_{\max}$, derived from the average link margin $\bar M_i$ of Eq.~\eqref{eq:margin} observed at location $i$ (locations with historically poor link margin score as higher risk, since an alert generated there is more likely to be delayed), and
    \item $\hat{G}_i$ is a static geospatial context score (e.g., terrain ruggedness, known storm frequency, or perimeter proximity to a road/waterway, sourced from external GIS/satellite-imagery layers rather than the node itself).
\end{itemize}
The weights $w_k$ are operator-configurable per deployment mode; a Sentinel deployment would typically weight $\hat B_i$ and $\hat G_i$ more heavily, while a Cargo route would weight $\hat I_i$ and $\hat C_i$ more heavily. Equation~\eqref{eq:cri} is, to our knowledge, not previously formulated in this specific combination; it is the paper's primary novel contribution and is intended as a starting hypothesis for empirical calibration rather than a validated model. The connectivity-risk term $\hat C_i$ specifically closes the loop between the communication-tier model of Section~\ref{sec:comm} and the geospatial layer, so that a route segment or fixed sentinel location with chronically poor GSM/LoRa/satellite margin is itself flagged as elevated risk, independent of whether an alert has yet failed to arrive there.

\subsection{Illustration}
\begin{figure}[htbp]
\centering
\includegraphics[width=0.78\textwidth]{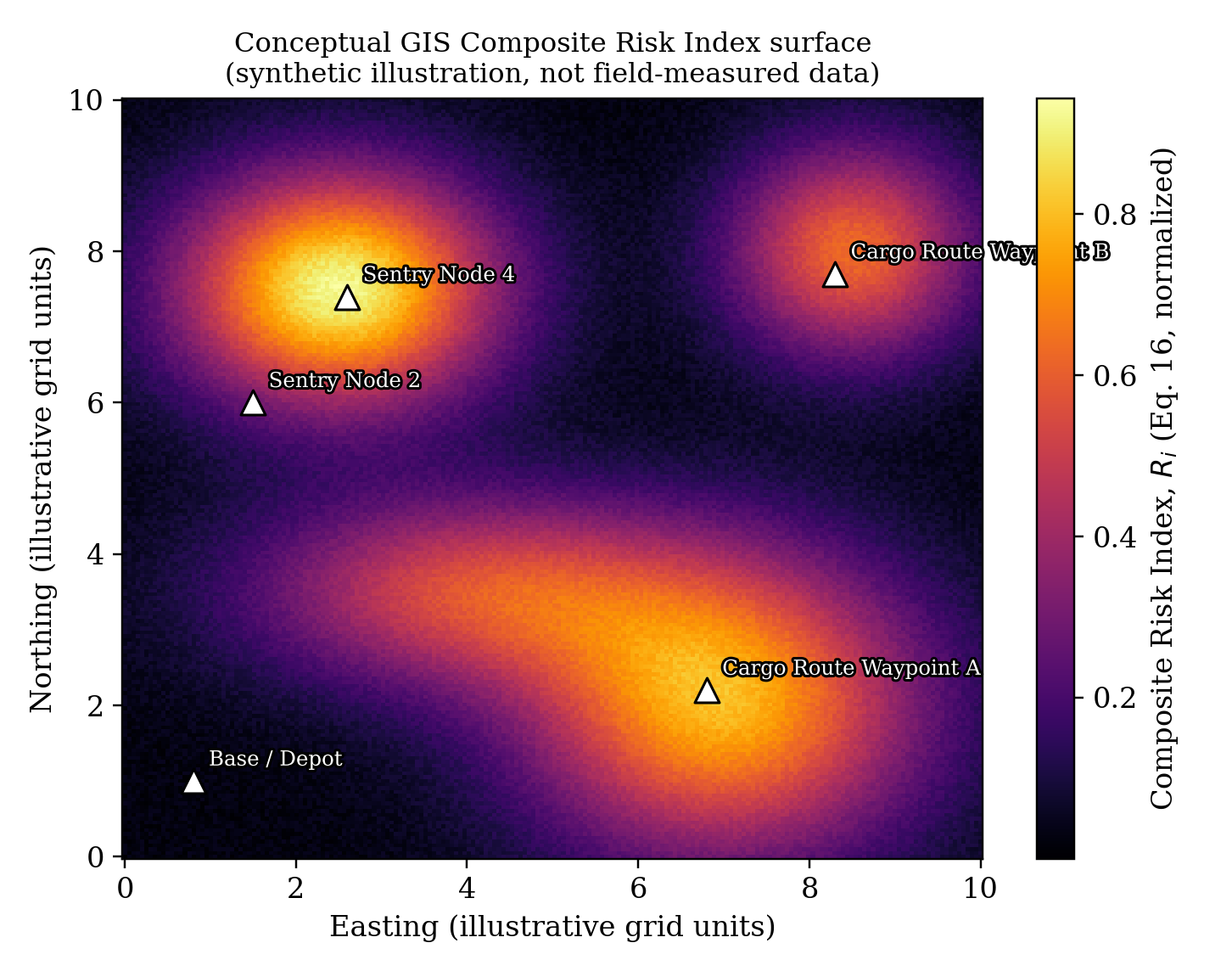}
\caption{Synthetic illustration of a Composite Risk Index surface (Eq.~\eqref{eq:cri}) over a shared operating area, with example Sentinel (fixed perimeter) and Cargo (route waypoint) node placements. Generated from a synthetic function for illustration only; not derived from field data.}
\label{fig:gis}
\end{figure}

Figure~\ref{fig:gis} shows a synthetic $R_i$ surface generated from Eq.~\eqref{eq:cri} with illustrative hotspot placements (a perimeter intrusion hotspot, a weak-connectivity corridor, a storm-prone maritime segment, and a rough-road segment), overlaid with example Sentinel and Cargo node locations, to illustrate how the two originally separate deployment modes populate a single shared risk map rather than two disconnected dashboards.

\section{Case Studies: Recovering the Two Original Designs}
\label{sec:case}

\subsection{Sentinel Mode -- Off-Grid Wildlife/Security Turret}
Setting \texttt{MODE = SENTINEL} recovers the original perimeter-sentry design: the ESP32 remains in deep sleep with only the PIR interrupt armed; on wake, the LiDAR range-gates the trigger against a stored background distance to suppress foliage/wind false alarms before committing to the active-tracking power budget; the fused IMU/LiDAR estimate (Section~\ref{sec:fusion}) drives the PID loop (Section~\ref{sec:pid}) to hold a camera, spotlight, or non-lethal deterrent on the detected target despite pole/branch sway; and an alert is dispatched via the tiered stack (Section~\ref{sec:comm}), defaulting to GSM with LoRa/satellite as fallback for sites with marginal cellular coverage -- addressing the single largest field-reliability gap of the original GSM-only design.

\subsection{Cargo Mode -- Active-Dampening Shipping Container}
Setting \texttt{MODE = CARGO} recovers the original cargo-protection design: the PIR sensor watches the crate's interior access point for breach events; the IMU continuously feeds the PID loop, which now holds a zero-tilt setpoint on a payload tray rather than tracking a target, with the same anti-windup-clamped control law of Eq.~\eqref{eq:pid}--\eqref{eq:antiwindup}; the LiDAR is repurposed to monitor internal clearance between payload and crate wall as an independent mechanical-limit check; and telemetry defaults to satellite-aware tier ordering, since maritime and cross-border routes cannot assume cellular coverage -- directly addressing the connectivity gap identified as a key limitation of the original GSM-only cargo design.

\section{Discussion and Limitations}
\label{sec:discussion}
This paper presents an architecture and its governing equations, not a field-validated system; the closed-loop response (Fig.~\ref{fig:pid}) is a simulation of the control model rather than a measurement from built hardware, and the communication trade-off (Fig.~\ref{fig:comm}) and risk surface (Fig.~\ref{fig:gis}) are literature-informed and synthetic illustrations, respectively. Open questions requiring empirical work include: (i) whether a single-point LiDAR provides sufficient false-alarm reduction in real foliage/weather conditions to justify its added power draw, (ii) the actual achievable weight-tuning of Eq.~\eqref{eq:cri} against ground-truth incident data, (iii) real-world LoRa gateway availability along representative cargo routes, and (iv) the mechanical feasibility of maintaining Eq.~\eqref{eq:antiwindup}'s travel limits under worst-case shock loading in the Cargo mode. Future work should instrument at least one physical prototype of each mode to collect the impact, breach, and connectivity data needed to calibrate the CRI weights empirically rather than by operator judgment alone.

\section{Conclusion}
\label{sec:conclusion}
We have shown that a perimeter/wildlife security turret and a high-value cargo protection crate can be expressed as two operating-mode instances of a single generalized IoT node architecture, sharing PIR-triggered wake logic, IMU/LiDAR sensor fusion, an anti-windup PID stabilization core, a tiered GSM/LoRa/satellite communication stack, and a shared GIS-based Composite Risk Index. This reframing surfaces connectivity and false-alarm limitations present in the original single-tier, PIR-only designs and proposes concrete, literature-grounded extensions -- LiDAR range-gating, LPWAN/satellite fallback tiers, RSSI-based local direction estimation, and a spatially-referenced risk index -- as a blueprint for a future unified hardware prototype.

\section*{Acknowledgments}
The author thanks colleagues at SmartData Technologies Ltd. and BRAC University for discussions that motivated the unification of these two originally separate embedded-systems designs.

\bibliographystyle{unsrtnat}
\bibliography{refs}

\end{document}